\documentstyle[12pt,aasms4]{article}

\newcommand {\ie} {{\it  i.e.}}

\newcommand {\eg} {{\it e.g.}}
\newcommand {\ea} {{\it et~al.}}

\newcommand {\be} {\begin{equation}}
\newcommand {\ee} {\end{equation}}

\begin{document}

\title{On Beaming Effects in Afterglow Light Curves}
\author{R.~Moderski$^{1,2,3}$, M.~Sikora$^{2,3}$, and
T.~Bulik$^{2}$}
\affil{$^1$ JILA, University of Colorado, Boulder \\
$^2$ Nicolaus Copernicus Astronomical Center, Warsaw \\
$^3$ ITP, University of California, Santa Barbara}

\begin{abstract}
The most luminous GRBs can be explained in terms of models involving
stellar mass central engines only if the ejecta are beamed.  As was pointed
out by Rhoads~(1997), the dynamics of the blast wave, formed by the beamed
ejecta sweeping the external gas, can be significantly modified by the
sideways expansion.  This is because in this case the surface of the blast
wave increases faster than just due to the radial divergence and so the
blast wave deceleration rate increases faster.  According to analytical
estimates, the effect becomes important shortly after the bulk Lorentz
factor of the blast wave drops below the inverse of the initial opening
angle of the beamed ejecta and is accompanied by a sharp break in the
afterglow light curve.

However, our numerical studies, which follow the dynamical evolution of the
blast wave, the evolution of the electron energy distribution, and take
into account the light travel effects related to the lateral size of the
source, show that the break of the light curve is weaker and much smoother
than the one analytically predicted. A prominent break emerges only for a
model without sideways expansion.
\end{abstract}


\section{INTRODUCTION}

Beaming of relativistic ejecta in GRBs has been postulated by many authors
in order to ease the GRB energy budget (see, \eg, M\'esz\'aros, Rees, \&
Wijers~1998 and refs. therein). There are basically two ways to verify the
beaming observationally: one is statistical and is based on counting the
afterglow like transient sources and comparing their rate with the GRB
rate, and the second one is related to the beaming effects predicted to be
imprinted in the afterglow light curves of individual objects
(Rhoads~1997).  Applying the first method to the X-ray transient sources,
Grindlay~(1999) found that results are consistent with no beaming
differentiating the GRB and X-transient rates.  However, as was pointed out
by Woods \& Loeb~(1999), the conclusive results about excess (or its lack)
of X-ray transients over GRBs must wait for much more sensitive future
instruments. This is because statistically significant contribution to the
excess of X-ray transients over GRBs is expected to be provided only by
weak X-ray transients, those representing afterglows phases when the bulk
Lorentz factor of the radiating shell drops below the inverse of its
angular size.  Similar studies can be performed also in optical and radio
band (Rhoads~1997; Woods \& Loeb~1999).

In individual objects, the beaming related effects are expected to be
imprinted in the optical and X-ray afterglow light-curves.  The lateral
expansion of the shocked, relativistic plasma causes that at some moment
the front of the blast wave starts to increase faster than due to the
cone-outflow (Rhoads~1997).  Due to this the blast wave begins to
decelerate faster than without the sideways outflows and this produces a
break in the light curve, the sooner the larger the beaming factor is.
Such a break is claimed to be present in the light curve of GRB~990123, the
most energetic GRB up to date (Kulkarni \ea~1999). Sari, Piran, \&
Halpern~(1999) speculate that afterglows with very steep light curves are
highly beamed. Possibly the break in such objects is not recorded because
it took place before the optical follow-ups.

As now, all theoretical studies of the light-curve breaks are analytical
and are based on: power-law approximation of the blast wave dynamics,
broken power-law approximation of radiation spectra and on ``on-axis''
relation between the observed flux and the emitted flux (Rhoads~1977, 1999;
Kulkarni \ea~1999; Sari, Piran, \& Halpern~1999).  In this paper, we treat
the dynamics using the prescription given by Blandford and
McKee~(1976). The evolution of radiation spectrum is calculated exactly, by
computing the time evolution of electrons from continuity equation and by
computing the observed luminosity through integrating the emitted radiation
over the ``$t= {\rm const}$'' surfaces. Our results show that the change of
the light curve slope is significant, but smaller than predicted
analytically. And, what is more important, the light curves steepen very
slowly, so that it is very difficult to talk about the specific time
location of the break. In order to demonstrate better the beaming effect,
we compare our results with the spherical case. We also show how the light
curve should look like if there is no lateral expansion.

In \S 2 we collect equations, which are used to compute the blast wave
speed, evolution of electrons, and afterglow light-curves.  In \S 3 we
present results of our numerical studies of afterglows produced by beamed
ejecta, and, in \S 4 we compare them with simple analytical estimations.

\section{BASIC EQUATIONS}

\subsection{Dynamics}

The deceleration of a blast wave is described by the following equations
(Blandford \& McKee~1976; Chiang \& Dermer~1998):
\be {d\Gamma \over dm} = - {\Gamma^2 -1 \over M} , \label{eq1} \ee
\be {dM \over dr} = {dm \over dr} [\Gamma - \epsilon_{rad}\epsilon_e
(\Gamma-1)] , \label{eq2} \ee
and
\be dm/dr = \Omega_j r^2 \rho = 2\pi r^2 (1-cos \theta_j) \rho ,
\label{eq3} \ee
where $\Gamma$ is the bulk Lorentz factor of the blast wave, $M$ is the
total mass including internal energy, $r$ is the distance from the central
engine to the blast wave, $dm$ is the rest mass swept up in the distance
$dr$, $\rho$ is the mass density of the external medium, $\epsilon_e$ is
the fraction of dissipated energy converted to relativistic electrons,
$\epsilon_{rad}$ is the fraction of electron energy which is radiated, and
$\theta_j$ is the angular size of the blast wave.  This angular size is not
constant but increases due to thermal expansion (Rhoads~1997),
\be \theta_j \equiv {a\over r} = \theta_{j0} + {v_l' \over c \Gamma},
\label{eq4} \ee
where the speed of the lateral expansion, $v_l'$, is assumed by
Rhoads~(1999) to be equal to the sound speed in the relativistic plasma,
$c_s = c / \sqrt{3}$, but considered by Sari \ea~(1999) to be relativistic.
Noting, that the plasma in the blast wave is continuously loaded by the
fresh gas which initially doesn't have any lateral bulk speed, one can
expect that in reality $v_l'$ does not reach relativistic value and sets up
somewhere between $c_s$ and $\beta_{\Gamma}c$, and in general depends on
$r$, and on $\theta_j$.
 
\subsection{Electron energy distribution}

We assume that the electrons are injected with the power low energy
distribution
\be Q = K \gamma^{-p}  , \label{eq5} \ee
with the minimum energy  of injected electrons
\be \gamma_m = {\epsilon_e (\Gamma-1) m_p \over m_e} \,{p-2 \over p-1}
. \label{eq6} \ee
The maximum energy of injected electrons for a given magnetic field, $B'$,
is assumed to be given by (de~Jager~\ea~1996):
\be \gamma_{max} \simeq 4 \times 10^7 {\left ( {B' \over 1{\rm G}} \right
)}^{-1/2} \ee

Normalization of the injection function, $K$, is provided by
\be L_{e,inj}' \equiv \int_{\gamma_m}^{\gamma_{max}} Q \gamma m_e c^2
~d\gamma = \epsilon_e {dE_{acc}' \over dt'} \, ,\label{eq7} \ee
where 
\be {dE_{acc}' \over dt'} = {dr \over dt'} {dE_{acc}' \over dr} = {dr \over
dt'} {dm \over dr} c^2 (\Gamma-1) = \Omega_j r^2 \rho \beta_{\Gamma} \Gamma
(\Gamma -1) c^3 , \label{eq8} \ee
is the rate of accreted kinetic energy, $dr = c \beta_{\Gamma} \Gamma dt'
$, $\beta_{\Gamma} = \sqrt {\Gamma^2 -1}/\Gamma $, and $t'$ is the time
measured in the blast wave comoving frame.

The evolution of the electron energy distribution is given by the
continuity equation
\be {\partial N_{\gamma} \over \partial r} = {\partial \over \partial
\gamma} \left(N_{\gamma} {d\gamma \over dr}\right) + Q , \label{eq9} \ee
where
\be {d\gamma \over dr} = - f(r)\gamma^2 - g{\gamma \over r} , \label{eq10}
\ee
are the electron energy losses. In both equations above, the derivatives
over comoving time $t'$ have been replaced by the derivatives over the
distance $r$, according to the relation $\partial /\partial t' = c
\beta_{\Gamma} \Gamma \partial /\partial r\, $. The first term on the rhs
of Eq.~(\ref{eq10}) represents synchrotron plus Compton energy losses, \ie,
\be f(r) = {\sigma_T \over 6 m_e c^2 } {B'^2 \over \beta_{\Gamma} \Gamma}
(1 + u_s'/u_B') , \label{eq11} \ee
where $u_B' = B'^2/8\pi$ is the magnetic energy density, and $u_s'$ is the
energy density of the synchrotron radiation, both as measured in the blast
wave frame.  The second term on the rhs of Eq.~(\ref{eq10}) represents the
adiabatic losses.  The parameter $g$ depends on the geometry of the
expansion; for 2-dimensional (lateral) expansion $g =2/3$, and for
3-dimensional expansion $g=1$.
 
We calculate the magnetic field following Chiang \& Dermer~(1999)
\be u_B' \equiv {{B'}^2 \over 8 \pi} = \epsilon_B \kappa \rho c^2 \Gamma^2 ,
\label{eq12} \ee
where $\kappa$ is the compression ratio and $\epsilon_B$ parameterizes the
departure of the magnetic field intensity from its equipartition value.

\subsection{Synchrotron  spectrum}

The evolution of the synchrotron spectrum in the blast wave frame is given
by
\be L_{syn,\nu'}'(r) = \int N_{\gamma}(r) P(\nu', \gamma) d\gamma ,
\label{eq13} \ee
where $P(\nu', \gamma)$ is the power spectrum of synchrotron radiation of a
single electron in isotropic magnetic field (see, \eg, Chiaberge \&
Ghisellini~1999).

The apparent monochromatic synchrotron luminosity as a function of time (a
light curve) is calculated from
\be L_{syn,\nu}(t, \theta_{obs}) = \int\!\!\!\int_{\Omega_j}
{L_{syn,\nu'}'[r(\tilde \theta)] {\cal D}^3 \over \Omega_j } d\cos {\tilde
\theta} d \tilde \phi , \label{eq14} \ee
where ${\cal D} = 1 /\Gamma(1-\beta_{\Gamma} \cos{\tilde\theta})$ is the
Doppler factor of the blast wave at the angle $\tilde\theta$.  The
coordinates ($\tilde\theta, \tilde \phi$) are chosen so that the observer
is located at $\tilde \theta=0$ ($\theta=\theta_{obs}$) and the jet axis is
at $\tilde \theta = \theta_{obs}$.  The integral is taken over the surfaces
\be t = \int {(1-\beta_{\Gamma} \cos \tilde\theta) \over c \beta_{\Gamma}}
\, dr = {\rm const} , \label{eq15} \ee
enclosed within the blast wave boundaries, $\Omega_j$.

\subsection{Inverse-Compton radiation}

We assume hereafter that cooling of relativistic electrons is dominated by
synchrotron radiation, \ie\ that $u_s' \ll u_B'$. This condition will be
verified and discussed in Appendix A.

\section{RESULTS}

We have used the following model parameters of the afterglow model in our
calculations: initial energy per solid angle, $E_0/\Omega_{j0} = 10^{54} \,
{\rm ergs \, s}^{-1} / 4 \pi$; $\Gamma_0 = 300$; $\theta_{j0} = 0.2$;
$\kappa = 4$; $\rho = m_p/1 {\rm cm}^3$; $\epsilon_e = 0.1$ (the
quasi-adiabatic case); $\epsilon_B = 0.03$; $p=2.4$.  The parameters were
not chosen to fit any specific observations, but rather to demonstrate the
difference between simple analytical predictions and self-consistent
numerical calculations regarding the beaming effects in a light-curves.

In Fig.~1 we present the dependence of a bulk Lorentz factor of the blast
wave on its distance from the central engine. We show three solutions for
three different values of $v_l'$: $v_l'=0$ (thin line); $v_l'=c/\sqrt {3}$
(solid line); and $v_l'=c$ (dotted line).  For $v_l'=0$ ($\to \theta_j={\rm
const}$) and $r_0 \ll r \ll r_{nr}$, the bulk Lorentz factor is well
approximated by
\be \Gamma \simeq \Gamma_0 \left(r_0/r \right)^{3/2} , \label{eq16} \ee
where 
\be r_0 \simeq \left( 3 E_0 \over \Gamma_0^2 \rho c^2 \Omega_j
\right)^{1/3} \simeq 1.2 \times 10^{17} {\rm cm} \,, \label{eq17} \ee
is the radius where deceleration of the GRB ejecta by sweeping of
interstellar gas starts to be efficient, and
\be r_{nr} \simeq r_0 \left (\Gamma_0\over 2 \right)^{2/3} \simeq 3.4
\times 10^{18} {\rm cm} \, , \label{eq18} \ee
is the radius above which the blast wave becomes nonrelativistic. Fig.~1
demonstrates that steepening of the $\Gamma(r)$ curves due to lateral
outflow is very smooth, without any sharp break like the one predicted
analytically to take place at a distance, at which $\Gamma$ drops below
$1/\theta_{j0}$, \ie\ at
\be r_D \simeq r_0 (\Gamma_0 \theta_{j0})^{2/3} \simeq 1.8 \times 10^{18}
{\rm cm} \, . \label{eq19} \ee

In Fig.~2 we show the radial dependence of the rate of kinetic energy
accreted by the blast wave, $ dE_{acc}'/ dt' $ (see Eq.~\ref{eq8}).  As one
could expect, the larger the lateral outflow speed, the larger the
accretion rate is. The steepening of curves at large $r$ is due to
transition from the relativistic regime ($\Gamma >2$) to the
nonrelativistic regime, where $dE_{acc}' /dt'$ is significantly reduced,
and becomes $\propto (\Gamma-1)$ (see Eq.~\ref{eq8}). When divided by
$\epsilon_e$, curves in Fig.~2 illustrate also the $r$ dependence of
injection luminosity of relativistic electrons (see Eq.~\ref{eq7}).

In Fig.~3 we show time evolution of the electron energy distribution,
$N_{\gamma}$, multiplied by $\gamma^2$. The curves are calculated at such
values of the radius $r$, from which the signal produced on the axis
$\tilde \theta=0$ is reached by the observer $t=1$,$ 10$, $10^2$, ...,
$10^7$ seconds after ``the signal'' from $r=0$. The relation between $r$
and $t$ is
\be t = \int_0^r {(1-\beta_{\Gamma})\over c \beta_{\Gamma}}\, dr \simeq
\int_0^r {1 \over 2 c \Gamma^2} \, dr . \label{eq20} \ee
The peak positions of the $N_{\gamma}\gamma^2$ curves mark Lorentz factor
of those electrons which carry most of leptonic energy at a given
distance. For injection spectral index $2 < p < 3$, the peak is located at
$\gamma_m$ given by Eq.~(\ref{eq6}), and this is the case in our
model. Another characteristic energy is
\be \gamma_c = 6.1 \times 10^{20} {m_p\over \epsilon_B \kappa \rho }
{1\over r \Gamma} , \label{eq21} \ee
below which the time scale of electron energy losses due to synchrotron
radiation is longer than the dynamical time scale. We present the
dependence of $\gamma_m$ and $\gamma_c$ on the radius $r$ in Fig.~4.

We can see from Fig.~4, that for the first 5 curves presented in the
Fig.~3, $\gamma_m > \gamma_c$. In this case, in accordance with the
analytical predictions, the electron spectra at $\gamma > \gamma_m$ are
well described by the power-law function, $N_{\gamma} \propto \gamma^{-s}$,
with the index $s=p+1$.  For $\gamma_c < \gamma < \gamma_m$, analytical
crude estimations predict $s=2$, which in our plot should be represented by
horizontal lines. This, however, is expected to be true only for $\gamma_c
\ll \gamma < \gamma_m$. In our model the ratio $\gamma_m/\gamma_c$ is not
large enough to provide space for $s=2$ and there is a smooth transition to
very hard low energy tail reached by electrons due to adiabatic losses.
For $\gamma_c > \gamma_m$, which is the case for the top 4 curves in the
Fig.~3, the predicted electron spectra should have a slope $s=p+1$ for
$\gamma> \gamma_c$, and $s=p$ for $\gamma_m < \gamma < \gamma_c$. The
former is seen, but the latter, again, due to narrow range between
$\gamma_c$ and $\gamma_m$ doesn't apply. Instead, there the log-energy
distribution is curved, smoothly joining the high energy portion of the
electron spectrum with its low energy adiabatic part. Let us note, that
very steep low energy tails of the curves on top of the plot result from
the fact that there is not enough time for electrons to drift adiabatically
to lower energies.  Note also, that the details of the low energy parts of
the electron energy distribution are not important, because contribution of
electrons from these parts to the observed radiation is negligible.

In Fig.~5 we present the observed radiation spectra computed for the same
sequence of $t$, as electron energy distributions shown in Fig.~3.  It
should be pointed out, however, that unlike in simple analytical
calculations, they are computed by integration of electron radiation from
$t={\rm const}$ surfaces (see Eq.~\ref{eq15}), \ie\ taking into account
light travel differences between photons emitted at different $\tilde
\theta$'s. The observed radiation spectra are peaked around $h \nu \sim
\Gamma \gamma_m^2 B/B_{cr} m_e c^2$, where rhs quantities are calculated
for $r$ given by Eq.~(\ref{eq15}) and $B_{cr} = 2 \pi m_e^2 c^3/h e \simeq
4.4 \times 10^{13}$ Gauss.  As we can see from Fig.~5, the high energy and
the low energy parts of the observed radiation spectra are well described
by power-law functions $L_{\nu} \propto \nu^{-\alpha}$, with $\alpha = p/2
= 1.1$ and $\alpha = -1/3$, respectively.  The former is produced, as
predicted analytically, by electrons with $\gamma >$ Max$[\gamma_m;
\gamma_c]$, the latter represents the low energy synchrotron radiation of
electrons with energies $\gamma <$ Min$[\gamma_c;\gamma_m]$. These high and
low energy spectrum portions are joined very smoothly without showing any
intermediate piece of the power-law spectrum.  This smoothing of the
observed radiation spectra results mainly from the fact that the observed
radiation at any given moment is contributed by radiation from $t={\rm
const}$ surfaces, \ie\ from different radii.

The light curves, computed for $\nu = 4.2 \times 10^{14}$Hz and $\nu = 2.5
\times 10^{17}$Hz, are shown in Fig.~6.  In this calculation we used $v_l'=
c/\sqrt{3}$ and considered two different locations of the observer, $\tilde
\theta=0$ and $\tilde \theta= 0.28$. The latter case is for the observer
located outside the initial ejecta cone.

In order to demonstrate better the beaming effect and its dependence on
$v_l'$, we plot in Fig.~7 four optical light curves; three for different
values of $v_l'$: $ 0$, $c/\sqrt{3}$, and $c$, and the fourth one for the
spherical outburst.  We can see, that for models with lateral expansion the
steepening of the light curves is extended over more than two time decades
and, down to nonrelativistic regime, doesn't reach analytically predicted
slope $\beta = p$ ($L_{\nu} \propto t^{-\beta}$). Sharp break is found only
for $v_l'=0$ model, and, it emerges shortly after $t(r_D)$, as
theoretically predicted.

We should note here, that our calculations of the model with the outflows
expansion are not fully consistent, because the Doppler factor includes
only the radial component of the bulk motion.  This, however, is expected
to affect only the results of the $v_l=c$ model, where we overestimate the
radiation contributed from the blast wave edge.

\section{DISCUSSION AND CONCLUSIONS}

The afterglows provide exceptional opportunity to study whether and how
much the GRB ejecta are beamed.  As predicted by Rhoads~(1997), the beamed
outflows should diverge from the cone geometry while decelerated by
sweeping up the external gas.  The sideways outflow of the shocked
relativistic plasma increases the front of the blast wave leading to a
faster deceleration.  Rhoads~(1997) showed, using simple analytical
analyses, that this should be imprinted in the light curve as a break
around $t(r_D)$, \ie\ when $\Gamma$ drops below $(c_l/c)/\theta$.  There
the light curve should steepen, changing the slope from $\beta =(3p-2)/4$
(for $\gamma > {\rm Max}[\gamma_c; \gamma_m]$) or from $\beta = 3(p-1)/4$
(for $\gamma_m < \gamma < \gamma_c$) (Sari, Piran, \& Narayan~1998) to
$\beta=p$ (Rhoads~1999).  Our numerical results agree only qualitatively
with these predictions.  The steepening does occur, however, the slope
change is smaller (to about $2.0$ instead of $p=2.4$) and is extended over
more than two decades of the observed time.

There are two reasons why, contrary to simple analytical estimations, the
distinct break does not emerge in our calculations.  First, as is shown in
Fig.~1, the dynamics of the blast wave is affected by the lateral outflow
very smoothly over the whole deceleration phase, and not just around $r_D$
(note, that at $r_D$ the blast wave area is already almost $4$ times larger
than it would be without the sideways expansion). Second, the observed
radiation at any given moment $t$ is contributed by the plasma which at
larger $\theta$ emits radiation from smaller $r$, and noting, that at
smaller $r$ plasma is moving faster and radiating stronger than at larger
$r$, the contribution of the off-axis plasma to the observed radiation is
larger than in the case of radiation contribution taken from $r={\rm
const}$ surfaces as calculated analytically.

The steepening of the light curve is predicted also for the beamed ejecta
without the lateral outflows. In this case, because no change of dynamics
and small light travel effects at $r >r_b$ (note that there the Doppler
cone becomes narrower than the ejecta cone), the break is very well
located, just around the time the $\Gamma$ drops below $1/\theta_j$, and
the light curve steepens by $\Delta \beta = 3/4$, in accordance with
analytical predictions (see, \eg, M\'esz\'aros \& Rees~1999).

It should be emphasized, however, that our treatment of the dynamics with
the sideways expansion is based on the approximation, that at any $r$ the
material is uniformly distributed across the blast wave.  In reality, the
lateral outflow can create $\theta $ dependent structure, with the density
of the swept material and the radial bulk Lorentz factor decreasing
sideways, and in this case the break in the light curve may become more
prominent.  2D-hydro relativistic simulations are required to verify this.

\acknowledgments

This project was partially supported by NSF grant No. AST-9529175, ITP/NSF
grant No. PHY94-07194, and the Polish KBN grant No. 2P03D 00415. MS and RM
thank Fellows of ITP/UCSB for hospitality during the visit and
participating in the program ``Black Hole Astrophysics''.  RM thanks NASA
for support under the Long Term Space Astrophysics grant NASA-NAG-6337.
MS acknowledges financial support of NASA/RXTE and ASTRO-E observing grants
to USRA (G. Madejski, PI).

\appendix

\section{INVERSE COMPTON COOLING}

The ratio of inverse Compton luminosity to synchrotron luminosity is given
by
\be { L_C' \over L_s'} ={ u_s' \over u_B' } , \label{A1} \ee
where
\be u_s'\simeq {L_s' \over 2 c \Omega_j r^2} , \label{A2} \ee
\be L_s' = (1- \eta_C) \epsilon_{rad} \epsilon_e {dE_{acc}'\over dt'} ,
\label{A3} \ee
$dE_{acc}'/ dt'$ and $u_B'$ are given by Eq.~(\ref{eq8}) and (\ref{eq12}),
respectively, and $\eta_C = L_C' / (L_C' + L_s')$.  Using all these
relations in Eq.~(\ref{A1}), we find that for $\Gamma \gg 1$
\be {L_C' \over L_s'} = {(1-\eta_C) \epsilon_{rad} \epsilon_e \over 2
\kappa \epsilon_B} , \label{A4} \ee
and noting that $L_C' /L_s' = \eta_C /(1 - \eta_C) $, we obtain
\be \eta_C = {(1 + 2\chi) - \sqrt {1 +4 \chi} \over 2 \chi} , \label{A5}
\ee
where
\be \chi = {\epsilon_{rad} \epsilon_e \over 2 \epsilon_B \kappa}
. \label{A6} \ee
and for $\chi \ll 1$, $\eta_C \simeq \chi$.

For $\gamma_m > \gamma_c $ practically all energy converted to electrons is
radiated ($\epsilon_{rad} \simeq 1$) and the inverse Compton is
energetically negligible ($\eta_C \ll 1/2$) if $\epsilon_B > 10^{-2}$. For
$\gamma_m < \gamma_c$, the luminosity peaks at $\nu_c$ and
\be\epsilon_{rad} \simeq \left(\gamma_m \over \gamma_c \right)^{p-2}
. \label{A7} \ee
In the latter case, the inverse Compton is energetically not important, if
\be {\gamma_c \over \gamma_m} \gg \left ( 10^{-2} \over \epsilon_B \right
)^{1 \over p-2} .  \label{A8} \ee
One can easily check, using the above criteria and Fig.~4, that for our
specific model the inverse Compton process does not dominate electron
cooling at any moment.

It should be noted, however, that the inverse Compton process can be
imprinted in the afterglow light curves, even if the Compton cooling is
less efficient than the synchrotron one.  At the moment when the Compton
component drifts down to the observed band, the light curve is expected to
flatten.  Chiang \& Dermer~(1999) demonstrated that this effect can be
visible in the X-ray light curves; in the optics it appears very late and
is already too weak to be observed, especially if overshined by the host
galaxy.

\clearpage

\begin{figure}

\centerline{\bf FIGURE CAPTIONS}

\caption{The bulk Lorentz factor of the blast wave as a function of a
distance from the central engine. {\it Thin line}: for $\theta_j = {\rm
const} = \theta_{j0}$; {\it solid line}: for $\theta_j = \theta_{j0} +
1/\sqrt 3 \Gamma$; {\it dotted line}: for $\theta_j = \theta_{j0} +
1/\Gamma$. The numbers along the curves show the observed time as measured
in seconds and given by Eq.~(\ref{eq20}).}

\caption{The accretion rate as a function of a distance from the center,
calculated for the same models as Fig.~1.}

\caption{Evolution of the energy distribution of relativistic electrons.
{From} the bottom to the top, the curves are for t: $1$, $10$, ..., $10^7$
seconds.}

\caption{The minimum injection electron energy, $\gamma_m$, and the
``cooling" electron energy, $\gamma_c$, as a function of time.}

\caption{The evolution of the apparent radiation spectra. The numbers mark
the observed times.}

\caption{The afterglow light curves: {\it the dotted lines} are the X-ray
light curves and {\it the solid lines} are the optical light curves. The
thick lines are for $\theta_{obs}=0$ while the thin lines are for
$\theta_{obs}=0.28$.}

\caption{The light curves in the time range, where the beaming effects are
strongest. {\it The thin solid line} is for $\theta_j = \theta_{j0}$; {\it
the thick solid line} is for $\theta_j = \theta_{j0} + 1/\sqrt 3 \Gamma$;
{\it the dotted line} is for $ \theta_j = \theta_{j0} + 1/ \Gamma$; and
{\it the dashed line} is for $\theta_j = \pi$ (spherical case).  }

\end{figure}

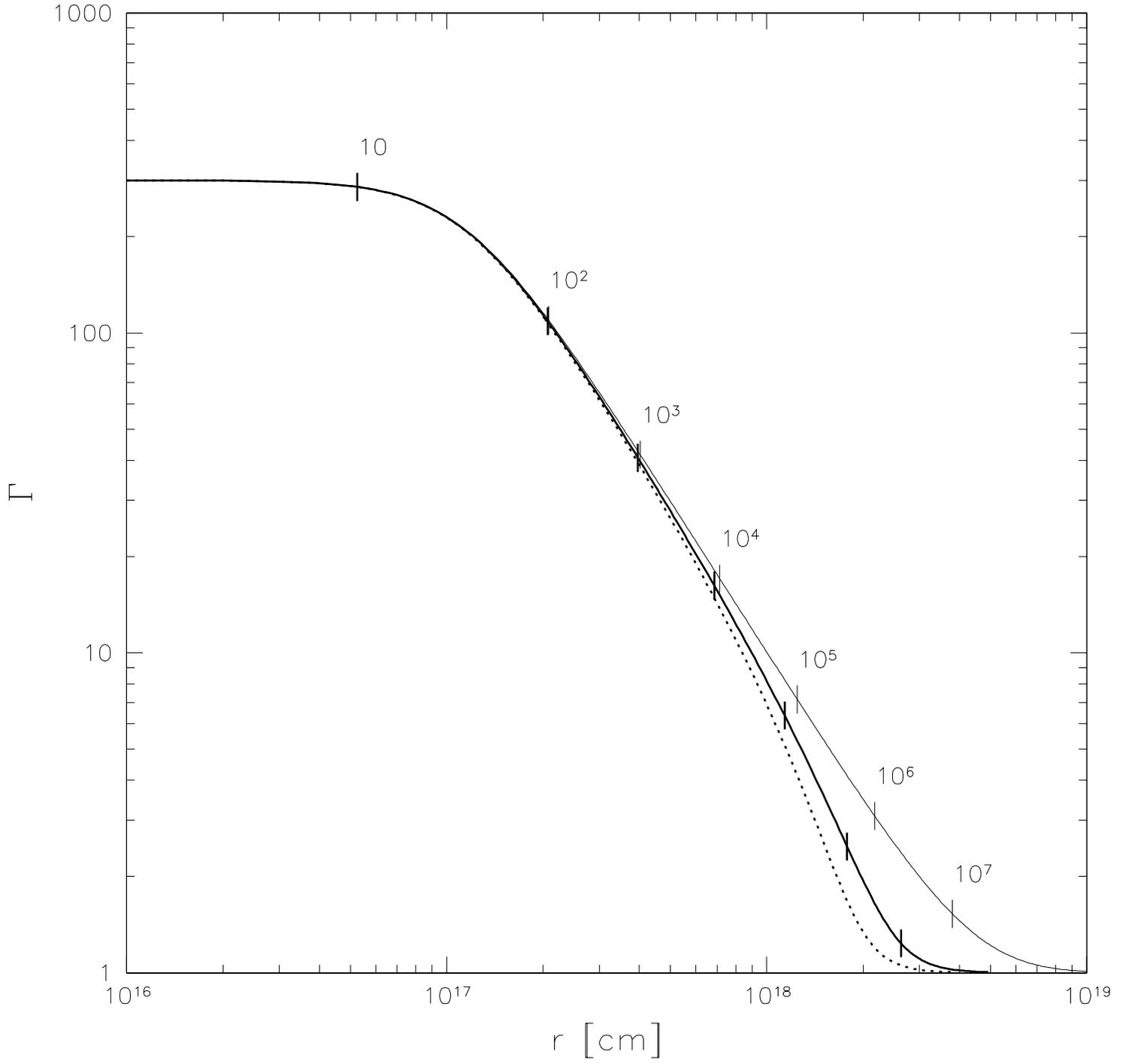
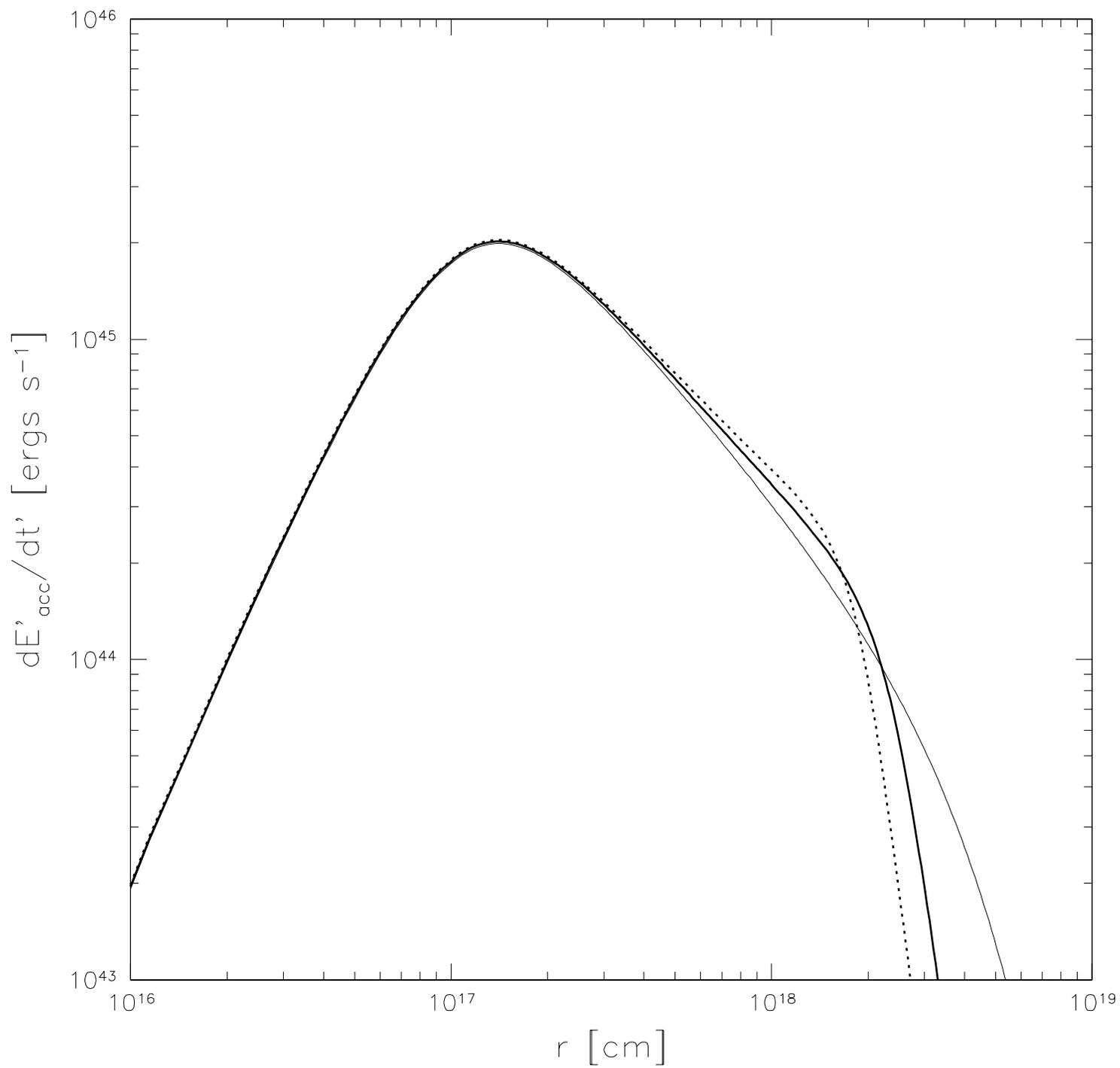
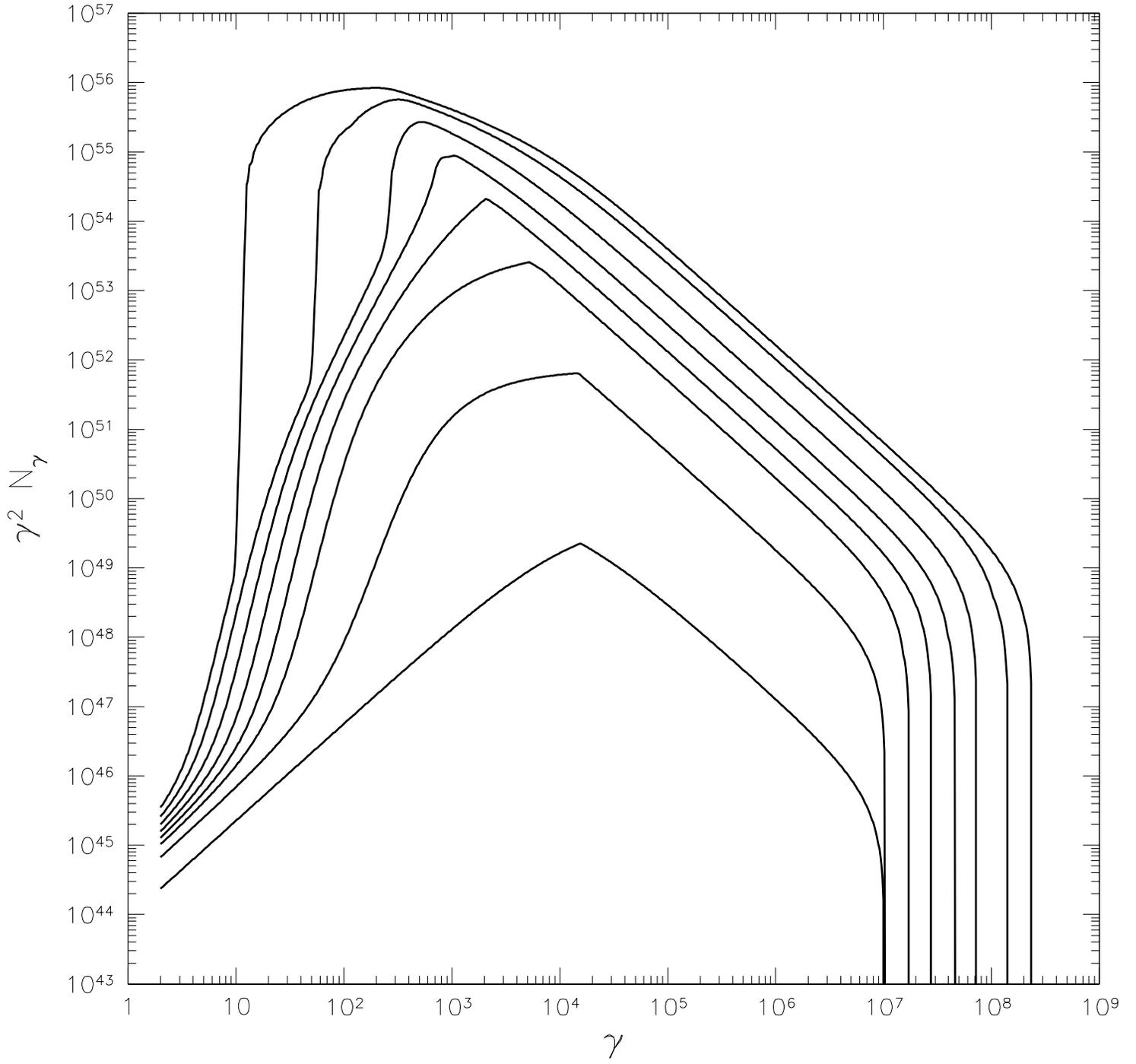
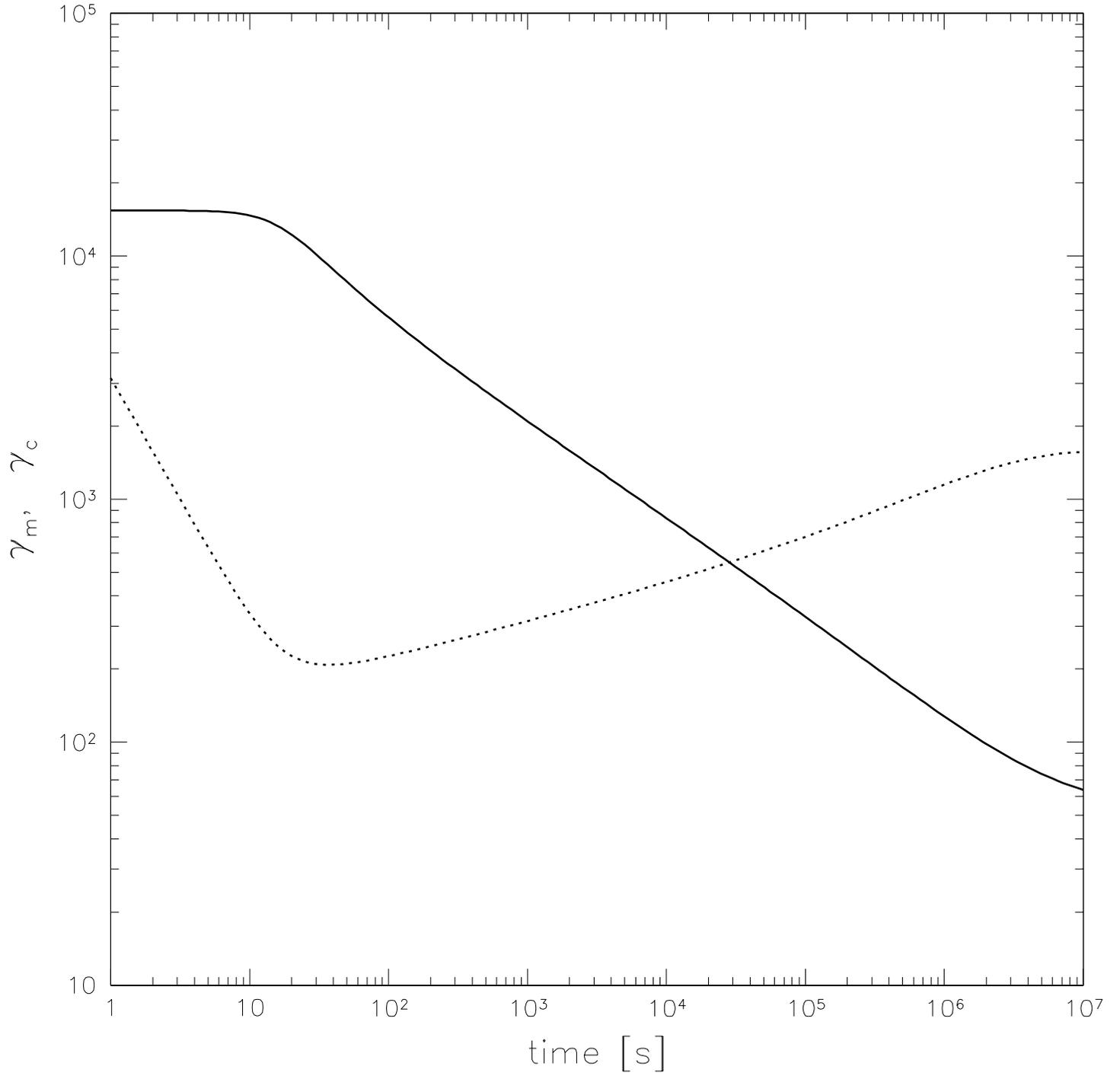
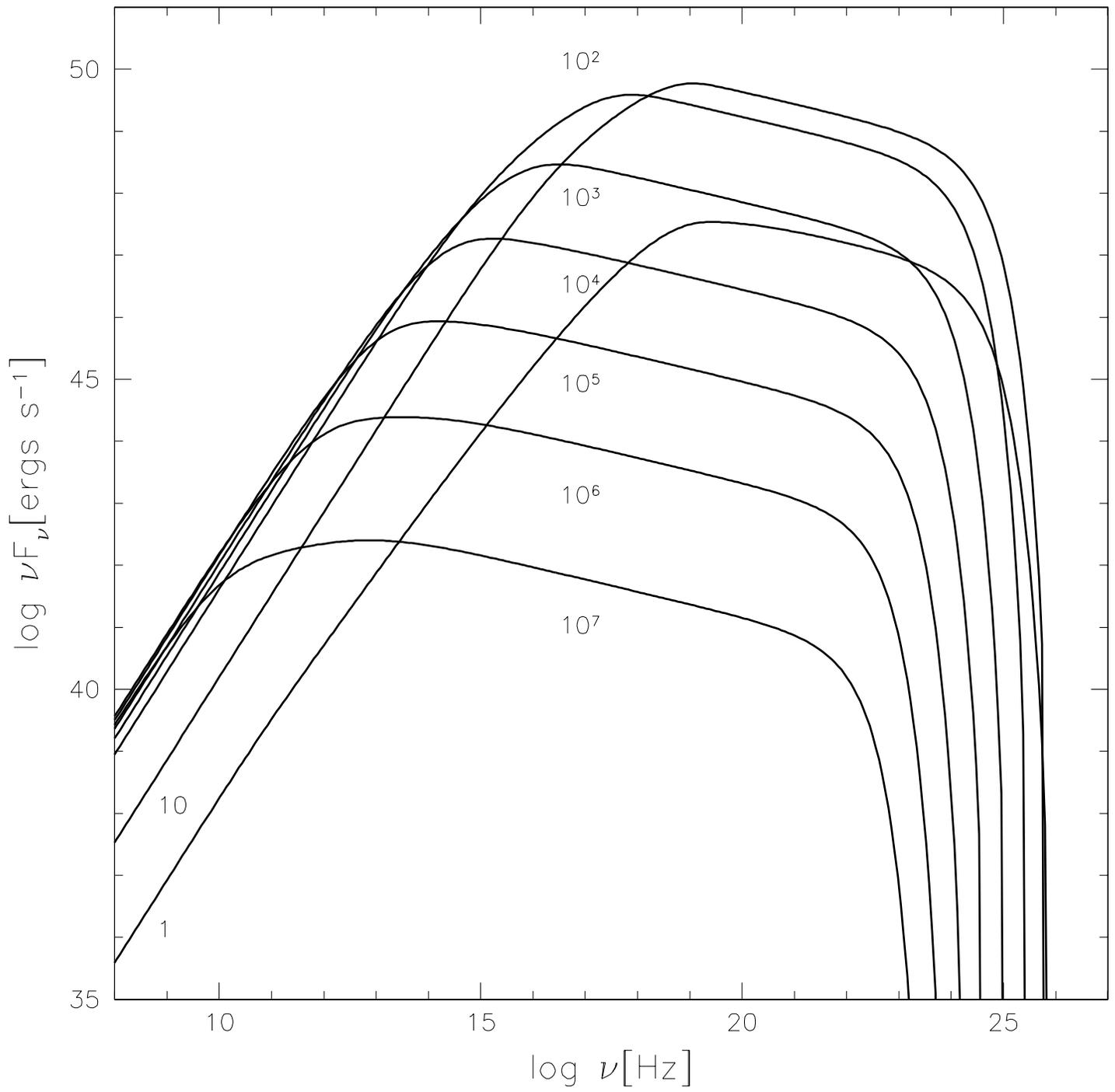
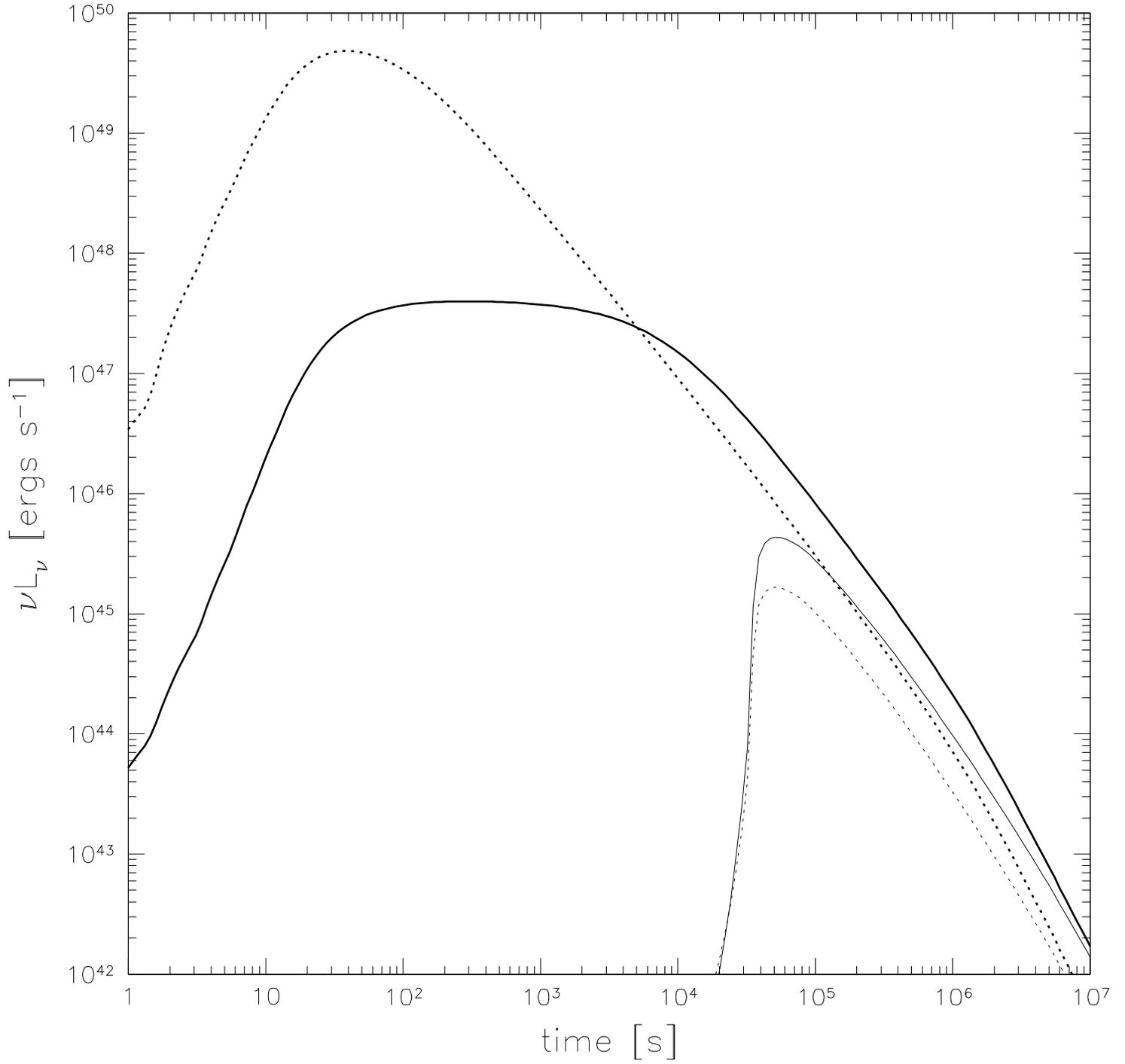
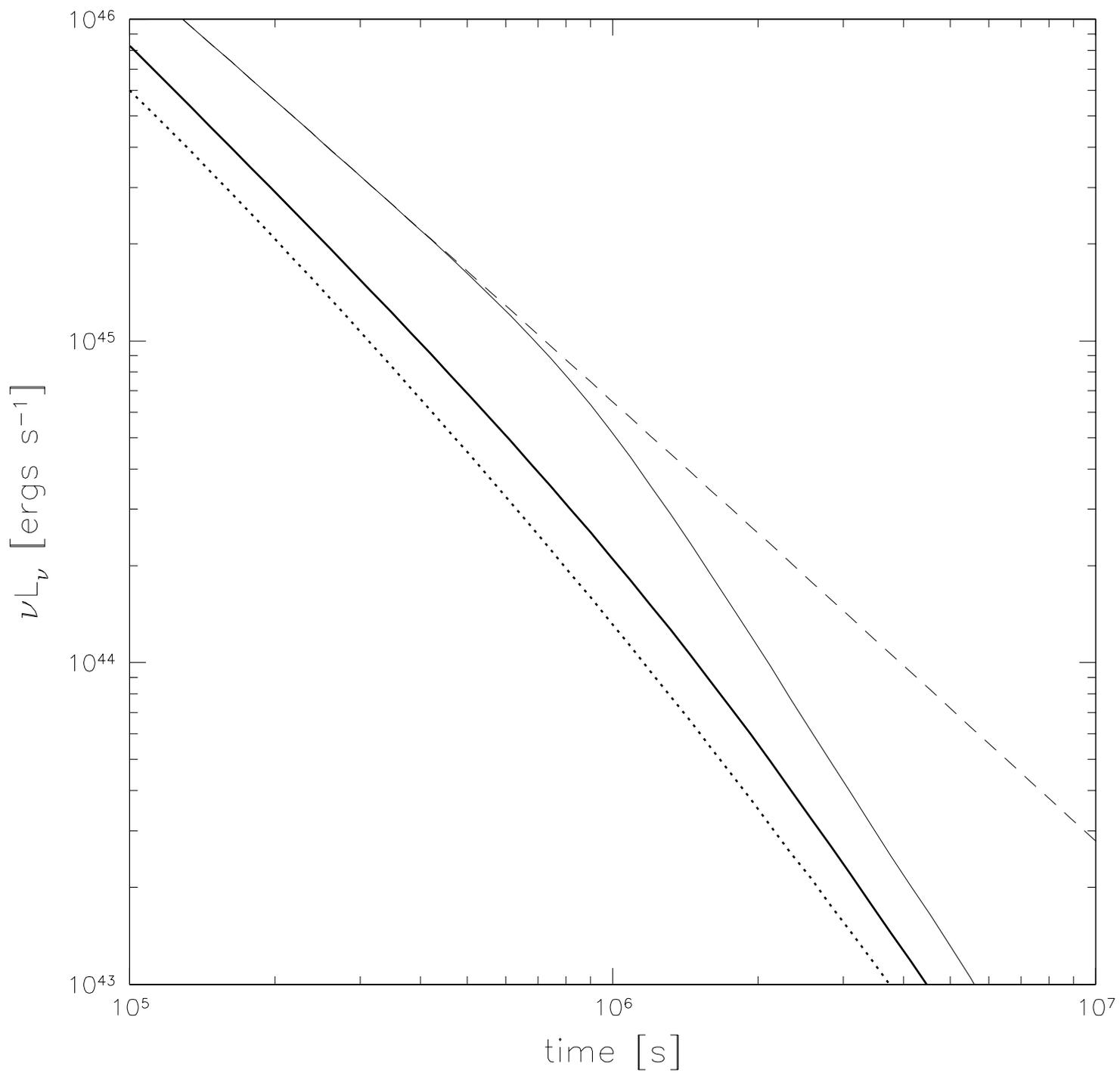
\end{document}